# Estimating linear radiance indicators from the zenith night sky brightness: on the Posch ratio for natural and light polluted skies

Salvador Bará[1,2,*], Xabier Pérez-Couto[2], Fabio Falchi[1,3], Miroslav Kocifaj[4,5], and Eduard Masana[6]

[1] *Departamento de Física Aplicada, Universidade de Santiago de Compostela, 15782 Santiago de Compostela, Galicia (Spain)*

[2] *A. Astronómica "Ío", A Coruña, Galicia (Spain)*

[3] *Istituto di Scienza e Tecnologia dell'Inquinamento Luminoso (ISTIL), 36016 Thiene, Italy.*

[4] *ICA, Slovak Academy of Sciences, Dúbravská cesta 9, 845 03 Bratislava, Slovakia*

[5] *Faculty of Mathematics, Physics, and Informatics, Comenius University, Mlynská dolina, 842 48 Bratislava, Slovakia*

[6] *Departament Física Quàntica i Astrofísica. Institut de Ciències del Cosmos (ICC-UB-IEEC), C Martí Franquès 1, E-08028 Barcelona, Spain*

* E-mail: salva.bara@usc.gal

**ABSTRACT**

Estimating the horizontal irradiance from measurements of the zenith night sky radiance is a useful operation for basic and applied studies in observatory site assessment, atmospheric optics and environmental sciences. The ratio between these two quantities, also known as Posch ratio, has been previously studied for some canonical cases and reported for a few observational sites. In this work we (a) generalize the Posch ratio concept, extending it to any pair of radiance-related linear indicators, (b) describe its main algebraic properties, and (c) provide analytical expressions and numerical evaluations for its three basic nighttime components (moonlight, starlight and other astrophysical light sources, and artificial light). We show that the horizontal irradiance (or any other linear radiance indicator) is generally correlated with the zenith radiance, enabling its estimation from zenith measurements if some a priori information on the atmospheric state is available.

ORCID

S. Bará            https://orcid.org/0000-0003-1274-8043

X. Pérez-Couto    https://orcid.org/0000-0001-5797-252X

F. Falchi         https://orcid.org/0000-0002-3706-5639

M. Kocifaj        https://orcid.org/0000-0001-9277-4692

E. Masana         https://orcid.org/0000-0002-4819-329X

# 1. INTRODUCTION

Estimating the value of the horizontal irradiance from measurements of the zenith night sky radiance is a very useful operation in several research fields, including astronomical site evaluation, light pollution monitoring, and environmental science. The interest in this kind of estimation has grown significantly in the last years, fostered by the progressive deployment of an increasing number of zenith sky brightness monitoring networks, promoted by goverment agencies as a part of their environmental data acquisition protocols (Bará 2016; Bertolo et al 2019), and by academic institutions for purposes of basic and applied research (Cavazzani et al 2020; den Outer et al 2015; Posch et al 2018; Pun & So 2012; Puschnig et al 2014; Zamorano et al 2016), as well as by many individual professional and amateur observers worldwide.

It is well-known from basic radiometric definitions that the horizontal irradiance $E$ produced by a hemisperic field of uniform radiance, $L$, is given by $E = \pi L$. Reversing propagation directions, this is also the relationship between $E$ and $L$ corresponding to a Lambertian emitter. For actual non-uniform radiance distributions, however, the ratio of the horizontal irradiance to the zenith radiance will in general differ from $\pi$ sr. The values of this ratio have been recently analyzed for several canonical cases by Kocifaj et al (2015), and observational data for some nights and sites have been reported by Jechow et al (2020). The latter authors proposed to name this quantity as the Posch ratio, PR, after one of the co-authors of the previous paper, Thomas Posch, who passed away at young age and is deeply missed by the light pollution research community.

The horizontal irradiance, however, is not the only useful indicator that can be derived from the radiance field. Other radiance-related indicators are appropriate for different kinds of studies, including the average all-sky radiance, the average radiance at intermediate zenith distances, or the average radiance in the first degrees of altitude above the horizon rim (Duriscoe, 2016; Falchi & Bará 2021).

In this paper we generalize the PR definition, extending it to any pair of radiance-related linear indicators. We describe its basic algebraic properties, as well as some of the physical processes that give rise to correlations between the values of the intervening indicators, providing that way the rationale for using the measured value of one of them to estimate the values of the others (Section 2). One of the basic properties of the PR in presence of several light sources is that it can be expressed as a linear combination of the values of the PR that each individual source would produce separately. Leveraging on that property, we provide in Section 3 numerical insights about the expected values of the Posch ratios produced by the three basic components of the night sky brightness (moonlight, starlight and other astrophysical light sources including terrestrial airglow, and artificial light). Discussion and conclusions can be found in sections 4 and 5, respectively.

## 2. THE POSCH RATIO FOR LINEAR SKY RADIANCE INDICATORS

By a linear radiance indicator is meant any physical quantity that can be obtained by the action of a linear operator on the spectral radiance of the sky, $L(z, \phi; \lambda)$, being $z$ the zenith distance, $\phi$ the azimuth and $\lambda$ the wavelength. Other variables on which the spectral radiance may depend (observer position, time, polarization state) are not indicated here, but should be implicitly kept in mind. The spectral radiance is usually reported in SI units $W \cdot m^{-2} \cdot sr^{-1} \cdot nm^{-1}$ or $photon \cdot s^{-1} \cdot m^{-2} \cdot sr^{-1} \cdot nm^{-1}$ and it corresponds, when expressed in the negative logarithmic scale of magnitudes per square arcsecond (mag/arcsec$^2$) within a given photometric band, to the astrophysical concept of 'surface brightness' (Bará et al 2020).

In this work we define a general Posch ratio as the ratio between the values of two such linear radiance indicators. A formal development of this definition, as well as some of the basic properties that can be derived from it can be found in the Appendix. For the purposes of this section we will focus on the Posch ratios defined as the quotient between two integral radiance indicators, of which the horizontal irradiance and the zenith radiance are particular instances.

An integral radiance indicator, $B_K$, has the following general form

$$B_K = \int_{z=0}^{\pi/2} \int_{\phi=0}^{2\pi} \int_{\lambda=0}^{\infty} K(z, \phi; \lambda)\, L(z, \phi; \lambda)\, d\lambda\, dz\, d\phi \qquad (1)$$

where $K(z, \phi; \lambda)$ is the kernel of the linear operator. Many useful indicators are defined by kernels that can be factored in an angular and a spectral part, such that $K(z, \phi; \lambda) = K_1(z, \phi) K_2(\lambda)$. For these indicators we have

$$B_K = \int_{z=0}^{\pi/2} \int_{\phi=0}^{2\pi} K_1(z, \phi)\, L(z, \phi)\, dz\, d\phi \qquad (2)$$

where

$$L(z, \phi) = \int_{\lambda=0}^{\infty} K_2(\lambda)\, L(z, \phi; \lambda)\, d\lambda \qquad (3)$$

is the sky radiance spectrally integrated within the photometric band $K_2(\lambda)$. It should be kept in mind that $L(z,\phi)$ depends on the choice of $K_2(\lambda)$ although, for the sake of simplicity, we will not indicate it explicitly in the notation.

Well-known radiance indicators described by Eq. (2) are the average upper hemisphere (all-sky) radiance, for which $K_1(z,\phi) = (2\pi)^{-1} \sin z$ for $0 \leq z \leq \pi/2$, the average radiance within a range of zenith distances $[z_1, z_2]$, for which $K_1(z,\phi) = \Omega_{12}^{-1} \sin z$ for $z_1 \leq z \leq z_2$ and zero otherwise, being $\Omega_{12}$ the solid angle (in sr) subtended by this band of zenith distances as seen from the observer, and the horizontal irradiance, for which $K_1(z,\phi) = (2\pi)^{-1} \cos z \sin z$. The metric factor $\sin z$ in these kernels combines with the differential linear angle elements $(dz, d\phi)$ of the observer spherical coordinate reference frame to provide the differential solid angles in the sky $d\Omega = \sin z \, dz \, d\phi$ (sr) required to perform the integrations over directions. Note that the zenith radiance, $L_0$, can be formally expressed as a particular instance of Eq. (2) by choosing the kernel as a Dirac-delta distribution centered at $z = 0$ and $\phi = 0$, i.e. by using $K_1(z, \phi) = \delta(z, \phi)$.

The potential usefulness of the Posch ratios $\Pi_{KK'} = B_K/B_{K'}$, and in particular of the ratios to the zenith radiance $\Pi_{K0} = B_K/L_0$, stems from the fact that the sky radiance $L(z,\phi)$ is expected to show some degree of angular correlation, due among other factors to the radiance mixing effects produced by the scattering of light in the atmosphere (see Appendix for details). Hence, $B_K$ and $L_0$ are expected to be correlated, and the values of $\Pi_{K0}$ are not expected to show an arbitrarily large variance. The radiance from any given direction of the sky generally consists of a direct component, propagated along geometrical rays from the sources located along the line-of-sight and attenuated by atmospheric extinction, plus a scattered component due to all sources present in the environment. The latter is composed of light scattered by the molecular and aerosol constituents of the atmosphere part of which is deflected into the line of sight. Hence, even in case that all radiating sources were spatially uncorrelated, different linear sky radiance indicators are expected to show some degree of correlation due to the scattered component. The purpose of this paper is to analyze whether these correlations are strong enough as to constrain the values of the Posch ratios $\Pi_{K0}$ within useful and known ranges, enabling that way their use as a predictive tool to estimate the (unknown) values of $B_K$ from the easily available measurements of the zenith radiance $L_0$.

An immediate result is that the Posch ratio of any actual sky whose radiance has contributions from several types of light sources can be expressed as a linear combination of the Posch ratios that each of these sources would produce separately. If the radiance $L(z,\phi)$ from the sky is the sum of the radiances $L_i(z,\phi)$ due to $i = 1, \ldots N$ source types, $L(z,\phi) = \sum_{i=1}^{N} L_i(z,\phi)$, it immediately follows that $L_0 = \sum_{i=1}^{N} L_{0,i}$ and $B_K = \sum_{i=1}^{N} B_{K,i}$, and hence $\Pi_{K0} = \left(\sum_{i=1}^{N} B_{K,i}\right) / \left(\sum_{i=1}^{N} L_{0,i}\right) = \sum_{i=1}^{N} \gamma_i \Pi_{K0,i}$ where $\Pi_{K0,i} = B_{K,i}/L_{0,i}$ is the $i$-th individual Posch ratio and $\gamma_i = L_{0,i}/L_0$ is the fractional contribution of the *i*-th source type to the total zenith radiance. Analog results hold for any other choice of the $\Pi_{KK'}$ pair or linear indicators. This suggests the convenience of performing a separate analysis of the PR produced by several kinds of skylight sources, whose details are addressed in Section 3 below.

## 3. POSCH RATIOS OF THE MAIN COMPONENTS OF THE NIGHT SKY RADIANCE

In the next subsections we describe the expected values of the $\Pi_{K0}$ ratio for the three main components of the night sky radiance: (i) starlight and other astrophysical sources including zodiacal light and terrestrial airglow, (ii) moonlight, and (iii) artificial light (light pollution). Some numerical examples are provided for the values of the PR of two particular indicators, the horizontal irradiance and the average upper hemisphere sky radiance, vs the zenith radiance, within the Johnson-Cousins V band (Bessell & Murphy 2012).

### 3.1. Posch ratios of the clear and moonless starry sky

The spectral radiance of the clear and moonless natural night sky is composed of the contributions of several types of extra-terrestrial terms, including starlight, zodiacal light, diffuse galactic light, extragalactic background, as well as of atmospheric airglow. The irradiance incident on the top of the atmosphere propagates through it along geometrical rays, being attenuated by absorption and scattering due to the molecular and aerosol atmospheric constituents. The radiance from any given direction of the sky is then composed of the direct, attenuated radiance of the sources present along the line of sight plus the radiance from other sources that is scattered into the field of view. The strength and effects of the scattering processess are strongly dependent on the particular state of the atmosphere at every moment of time.

The natural night sky radiance has been measured and modeled since long time ago in different spectral regions, a work that continues up to our days (Roach & Gordon 1973; Turnrose 1974, Benn & Ellison 1998, Leinert et al 1998, Patat 2008, Noll 2012, Duriscoe 2013). A multi-band model with an updated estimate of the integrated starlight term based on the DR2 Gaia catalogue has been recently released by Masana et al (2020), with an updated version based on the EDR3 dataset already available online (GAMBONS https://gambons.fqa.ub.edu/). The GAMBONS model includes the above mentioned types of sources, providing angularly resolved multiband radiance estimations across the sky vault, as well as the values of several radiance-related indicators like the zenith radiance (average radiance within 5° of the local zenith), the average upper hemisphere radiance, and the horizontal irradiance, given in energy units or photon number according to the original definitions of the specific photometric bands.

The different terms contributing to the sky radiance are best expressed using different spherical reference frames. Whereas the integrated starlight and the diffuse galactic light are more naturally described in galactic coordinates, zodiacal light is canonically described using ecliptic coordinates and is strongly dependent on solar elongation. The mean, time-averaged spatial component of the airglow, in turn, has a simpler expression in local alt-azimuth coordinates, showing a continuous gradient from zenith to horizon described by the van Rhijn function (Leinert et al 1998), with an azimuthally symmetric distribution.

The Posch ratio for the horizontal irradiance vs the zenith radiance is expected to show seasonal variations and also faster changes within a single night, due to the different regions of the celestial sphere located above the observer's horizon, to the atmospheric and airglow variability, and to the particular celestial features located in the zenith region at every instant. The same can be said of the Posch ratio for the average radiance of the upper hemisphere. To get some insights about the time evolution and the expected values of these ratios, we have computed them using the GAMBONS model in the Johnson-Cousins V band for an observer located at sea-level at 40.0° N 0.0°E, one out of every ten successive nights at 00:00 UTC, spanning the whole year. Atmospheric absorption and scattering were accounted for by using the atmospheric point-spread function of Kocifaj and Kránicz (2011), with a scattering phase function composed of molecular (Rayleigh) and aerosol (Henyey–Greenstein) components, weighted by the corresponding optical depths (Masana et al 2020). Additional parameters were the aerosol single-scattering albedo 0.85, Ångström exponent 1.0, airglow intensity 100% of nominal in GAMBONS, and a variable range of aerosol optical depths (AOD, 0.05 to 0.40) and asymmetry parameters ($g$, 0.3 to 0.9). The asymmetry parameter is the mean over the sphere of the cosine of the scattering angle weighted by the scattering phase function (Irvine 1963, Mischenko et al 2002), and informs about whether the scattering is predominantly forward (*g>0*) or backward (*g<0*) directed.

Figure 1 shows the seasonal evolution of the V-band zenith radiance, the average radiance of the upper hemisphere and the horizontal irradiance for an atmosphere with AOD=0.20 (at $\lambda$=550 nm) and aerosol asymmetry parameter *g*=0.6. The oscillation of the zenith radiance and horizontal irradiance are fundamentally driven by the presence of the Milky Way high in the sky during winter and summer, the latter being brighter.

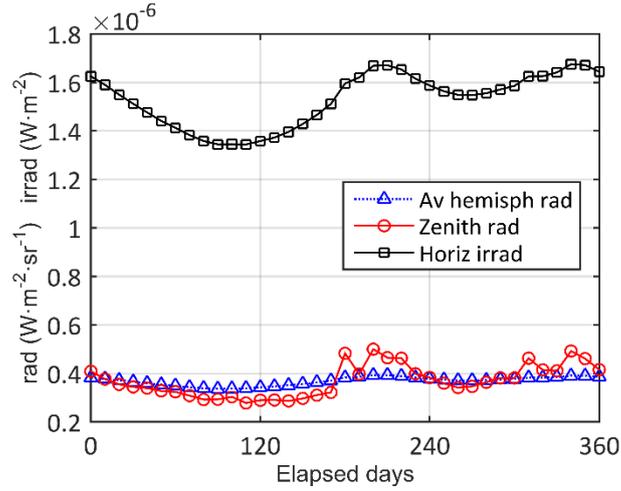

**Figure 1**. Zenith sky radiance $L_0$, average hemispheric radiance $L_a$, and horizontal irradiance $E_0$ of the clear and moonless natural sky in the Johnson-Cousins V band, computed for an observer at 40.0° N, 0.0°E at local midnight (UTC) along the year (see text for details).

The corresponding Posch ratios are shown in Figure 2, for different combinations of the atmospheric parameters. It can be seen in this Figure (left column) that the Posch ratio for the horizontal irradiance spans the range 3-5 sr (recall that this particular Posch ratio is not a dimensionless parameter). These values are of the same order of magnitude but slightly larger than the one corresponding to a uniformly bright hemisphere, $\pi$ sr. This is an expected result, since the average hemispheric radiance of the starry sky tends to be larger than the zenith one (Figure 2, right column). The presence of the Milky Way at high angular elevations over the horizon increases significantly the zenith radiance, producing pronounced dips of the Posch ratios in winter and summer. The behaviour of all curves is very similar, with the Posch ratios decreasing for thicker atmospheres (larger AOD) and increasing slightly for larger aerosol asymmetry parameters (more forward scattering, and hence less scattered zenith radiance everything else being equal). Overall, this Figure shows that there is a reasonably deterministic evolution of the Posch ratio of the clear and moonless natural sky across the year (excepting for the highly variable airglow contributions, that can significantly distort this picture). In absence of information about the state of the atmosphere the values of the horizontal irradiance and the average upper hemisphere radiance can be determined within a range of ~3 to ~5 sr times the zenith radiance for the former and ~0.7 to ~1.4 (dimensionless) times for the latter. The estimation can be made substantially more precise if some atmospheric information (AOD and parameter *g* of the aerosols) is available, either from real-time measurements or from recorded long-term site statistics.

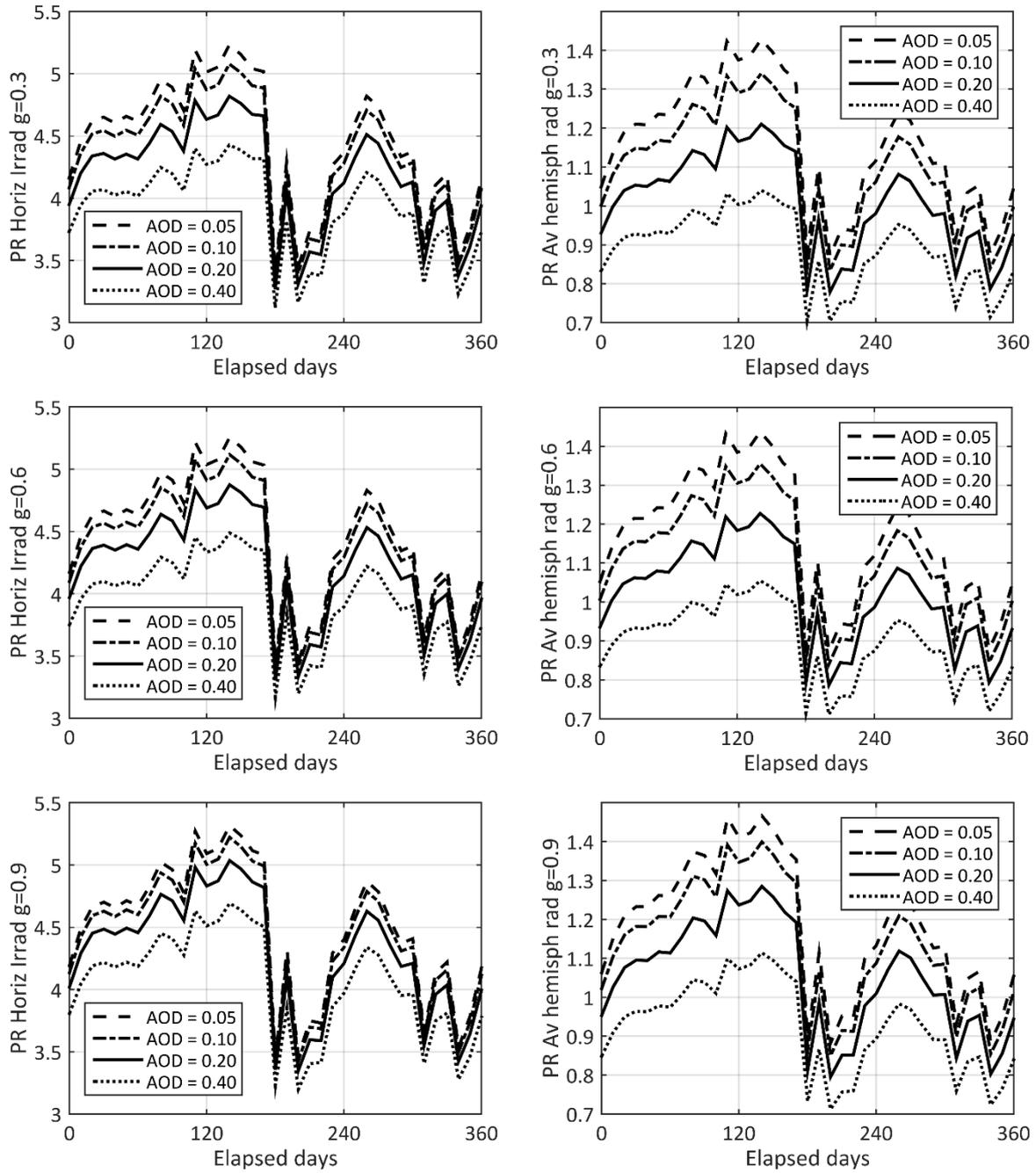

**Figure 2**. Posch ratios of the horizontal irradiance $E_0$ (left column, units sr), and of the average upper hemisphere radiance $L_a$ (right column, dimensionless) of a clear and moonless natural sky, versus the average radiance in a region of radius 5° around the local zenith, $L_0$, for an observer in the conditions described in the text and for three values of the aerosol asymmetry parameter, *g*: 0.3 (upper row), 0.6 (middle), 0.9 (bottom).

## 3.2. Posch ratios of moonlight

The radiance of the night sky due to the Moon, $L_M(z, \phi; \lambda)$, can be expressed as the sum of two terms: the direct radiance $L_D(z_M, \phi_M; \lambda)$, from the direction $(z_M, \phi_M)$ the Moon occupies at each moment in the sky, and the scattered moonlight radiance, $L_S(z, \phi; z_M, \phi_M; \lambda)$, arriving to the observer from all sky directions $(z, \phi)$, including the light scattered forward along the Moon-observer line of sight.

The direct term can be easily calculated as a spectrally attenuated version of the extra-atmospheric Moon radiance, according to the particular state of the atmosphere and the airmasses travelled by the geometrical light rays. The scattered term can be calculated from the extra-atmospheric spectral radiance of the Moon by using the Kocifaj-Kránicz (2011) model for the spectral radiance of the sky, applied here to the Moon, not to the Sun as in the original authors' work. These type of sky models, based on a two-parameter continuous function defined on the upper hemisphere (Kocifaj 2009), provide a rigorous physical basis to other commonly used daylight sky radiance distributions (Kittler 1998; Kocifaj 2013; Kittler & Darula 2021; Kocifaj & Kómar 2021), and have also proven to be particularly well adapted to compute the sky radiance due to ground-based artificial light sources (Kocifaj & Bará 2019).

For the purposes of this work we have used the spectral atmospheric point spread function described by eq.(18) of Kocifaj and Kránicz (2011), acting on the extra-atmospheric spectral Moon irradiance. The spectral Moon irradiance was calculated in turn using the STIS002 spectral irradiance of the Sun from the CALSPEC library (Bohlin, Gordon & Tremblay 2014), and the ROLO model for the lunar albedo (Kiefer & Stone 2005), interpolated in wavelengths and Moon phase angles, smoothed, and with corrections by Velikodsky et al (2011), Jones et al (2013), and Román, et al (2020). The numerical examples shown below were computed for the full Moon (phase angle 0°), with a Sun-Earth distance 1 au, and a Moon-Earth distance of 384400 km. The atmospheric Kocifaj-Kránicz distribution was applied to an exponential stratified atmosphere with different values of AOD (at λ=550 nm) and Henyey-Greenstein asymmetry parameters (*g*), an Ångström exponent 1.0, aerosol albedo 0.85, molecular optical depth $\tau_R = 0.00879\, \lambda^{-4.09}$, with $\lambda$ in micron (Teillet 1990), and an airmass function $M(z) = 1/[\cos z + 0.50572 \cdot (96.07995 - z)^{-1.6364}]$, with $z$ in degrees (Kasten & Young 1989), parameters also used in the GAMBONS calculations (see section 3.1). As in the previous subsection, results are provided here for the Johnson-Cousins V band, and the zenith radiance corresponds to the average radiance in a region of radius 5° around the zenith. The direct Moon radiance has been taken into account for the computation of the zenith radiance in proportion to the Moon solid angle fraction contained within the 5° zenit circle.

The resulting Posch ratios for the horizontal irradiance and the average upper hemisphere radiance are shown in Figure 3. Again, larger AOD values translate into smaller Posch ratios for both radiometric quantities, and larger values of the asymmetry parameter translate into larger values of the ratios for most zenith angles of the Moon, due, as in the subsection above, to the relatively smaller zenith scattered radiance. This last result however does not hold when the Moon is within the 5° zenith circle, since in that case the direct radiance of the Moon and stronger forward-peaked aerosols substantially increase the denominator of the ratios. Independently from the value of $g$, the direct Moon radiance in the zenith region dramatically reduces the ratios, which become of order 0.032 rad for the horizontal irradiance and 0.006 (dimensionless) for the average radiance of the upper hemisphere, both averaged over the twelve atmospheric conditions shown in the figure (4 AOD x 3 $g$). When the Moon is close to the horizon the direct radiance is strongly attenuated, and the horizontal irradiance is further reduced by the low value of the cos $z$ term. In these conditions the Posch ratios take again values of order of magnitude comparable to those of the clear and moonless starry sky, namely 7.53 rad for the horizontal irradiance and 5.40 for the average upper hemisphere radiance, averaged over the twelve atmospheric conditions here considered.

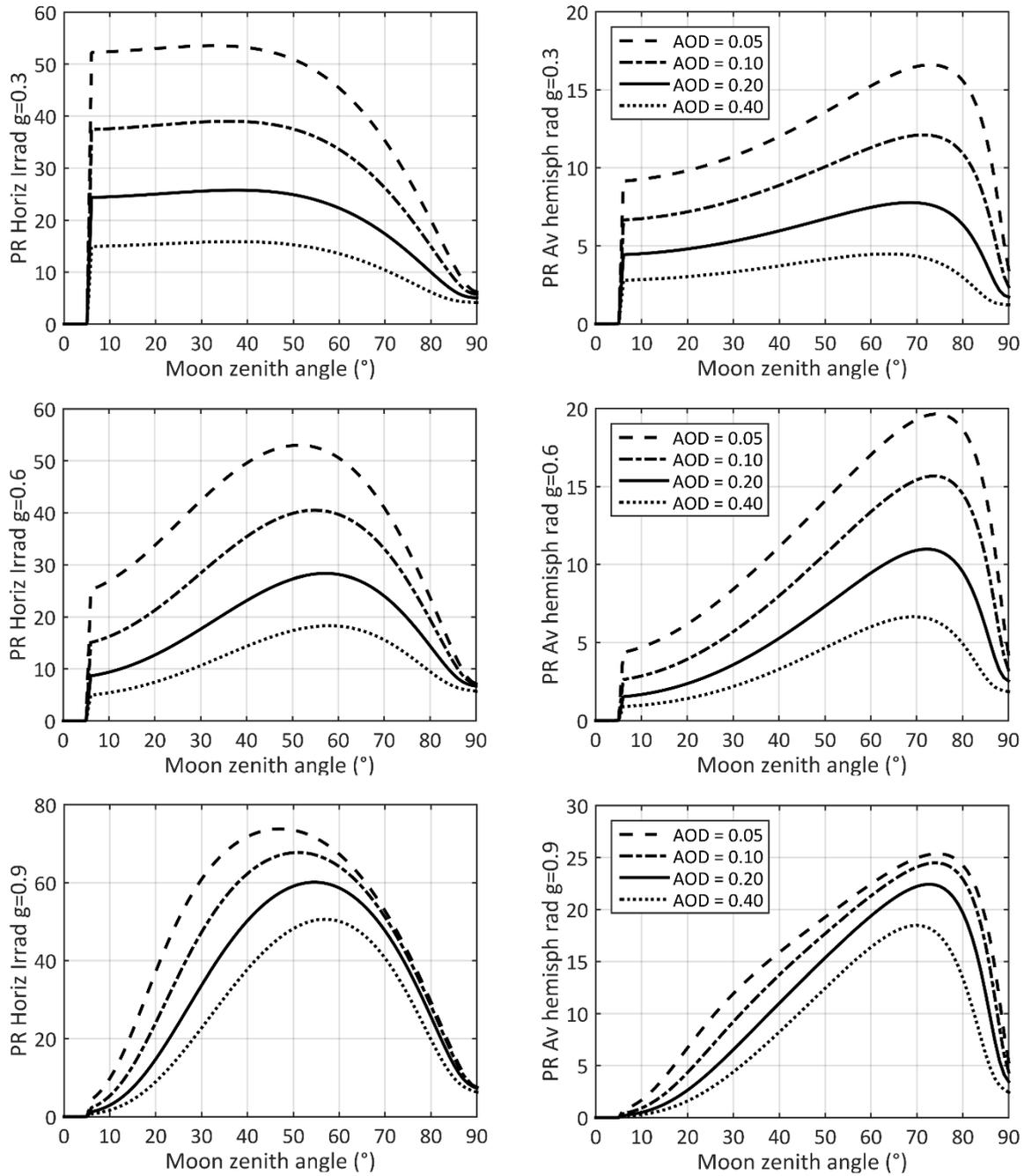

**Figure 3**. Posch ratios for the horizontal irradiance $E_0$ (left column, in sr), and for the average upper hemisphere radiance $L_a$ (right column, dimensionless) of a moonlit sky, versus the average radiance in a region of radius 5° around the local zenith, $L_0$ (see text for details)

### 3.3. Posch ratios of the light pollution component of the night sky brightness

The night sky radiance due to artificial lights can be computed using suitable light propagation functions, dependent on the spectral and angular emission patterns of the sources and on the state of the atmosphere. Several analytical and numerical approaches are nowadays available (see e.g. Garstang 1989; Cinzano & Falchi 2012; Kocifaj 2007, 2018; Aubé & Simoneau 2018; Bará et al 2019; Aubé et al 2020; Simoneau et al 2021), enabling the calculation of this radiance within any predefined photometric band, and, subsequently, the value of any desired linear radiance indicator, within the range of validity of each approach.

As shown in Duriscoe et al (2018), Bará & Lima (2018), and Falchi & Bará (2020), the linear light pollution indicators can be directly calculated from the source emissions, without the need of performing the intermediate calculation of the angularly resolved sky radiance. The problem is then formulated as a two-dimensional linear systems operation whereby the source emissions are transformed into the final indicators by means of a superposition integral whose kernel is the specific point spread function corresponding to the indicator under study (Bará et al 2021; Falchi & Bará 2021). In case the point-spread function is invariant under translations, the superposition integrals become two-dimensional convolutions and can be efficiently evaluated in the Fourier domain (Bará et al, 2020).

The Posch ratio of the $B_K$ indicator with respect to the zenith radiance can be calculated from the spatial distribution of the emitted source radiances $L_1(\mathbf{r}')$ as

$$\Pi_{K0} = \frac{B_K(\mathbf{r})}{L_0(\mathbf{r})} = \frac{\int_{A'} K(\mathbf{r},\mathbf{r}') L_1(\mathbf{r}') \, d^2\mathbf{r}'}{\int_{A'} K_0(\mathbf{r},\mathbf{r}') L_1(\mathbf{r}') \, d^2\mathbf{r}'} \qquad (4)$$

(see Appendix), where $\mathbf{r}$ and $\mathbf{r}'$ are the position vectors of the observer and the sources, respectively, $K(\mathbf{r},\mathbf{r}')$ is the point spread function describing the propagation of the indicator produced by an elementary light source of radiant intensity $L_1(\mathbf{r}') \, d^2\mathbf{r}'$, and $K_0(\mathbf{r},\mathbf{r}')$ is the point spread function for the propagation of the zenith radiance.

To get some insights about the expected values of the Posch ratios associated with the horizontal irradiance and the average upper hemisphere radiance due to artificial lights we have used the results from Falchi & Bará (2021) encompassing the Iberian peninsula, the Balearic islands, and a region of Northern Maghreb (Figure 4). In that work several radiance-related indicator maps were produced for a wide territory by using point spread functions previously calculated as a function of the distance of the observer to an artificial light source of unit amplitude. To that end, hemispheric (all-sky) radiance distributions were first calculated by using a Garstang-Cinzano model of artificial light propagation in the atmosphere for a spherical Earth, evaluating the radiance at more than one hundred thousand directions of the sky uniformly distributed within a Zenith Equal Area projection of the hemisphere above the observer. These hemispheric radiance distributions were computed for observers located from 0.12 to 527 km away from the artificial source, and for different atmospheric conditions. The point spread function for each individual indicator (zenith radiance, horizontal irradiance...) was subsequently calculated by angularly integrating the hemispheric radiance weighted by the function that defines the indicator. These point spread functions, in combination with satellite nighttime lights imagery, allow to compute the value of the indicators at any observer position by adding up the contributions of all artificial light sources in the surrounding territory. Note that although the formulation in Eq. (4) is totally general and can be applied to any type of point spread function, the functions used in Falchi & Bará (2021) were calculated for the average altitudes above sea level of the sources and the observers, respectively, resulting in a set of shift-invariant, rotationally symmetric PSFs that allowed the efficient production of the maps using fast Fourier techniques. The numerical results presented below shall be assessed taking into account this simplification.

The scatter plots versus the zenith radiance and the histograms of the Posch ratios for all pixels of the emerged lands are shown in Figure 5. Table 1 provides the coefficients of the second-degree polynomials fitted to the scatter plots. These regression equations were subsequently used to predict the values of the horizontal irradiances and average upper hemisphere radiances in a region of Northern Africa located immediately eastward of the displayed maps, containing 177559 pixels, resulting in relative estimation errors of 0.074 (std 0.046) and 0.109 (std 0.081), respectively.

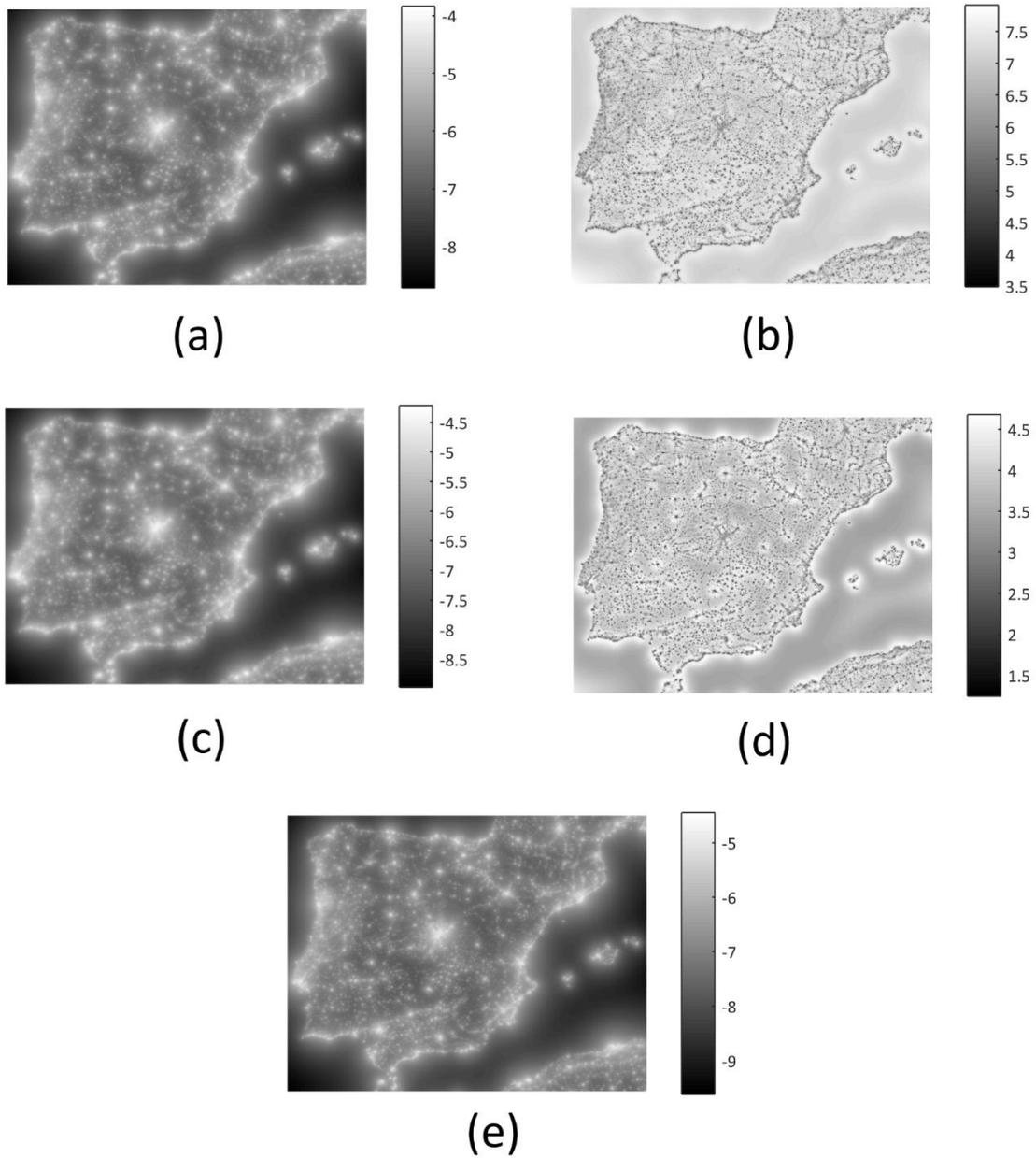

**Figure 4**. Posch ratios of the artificial brightness of the night sky in the Johnson V band for the Iberian peninsula, Balearic islands, and a region of Northern Maghreb. (a) $\log_{10}(E_0)$, being $E_0$ the horizontal irradiance in W·m$^{-2}$; (b) Posch ratio for the horizontal irradiance $E_0/L_0$, in sr; (c) $\log_{10}(L_a)$, being $L_a$ the average upper hemisphere radiance in W·m$^{-2}$·sr$^{-1}$; (d) Posch ratio for the average upper hemisphere radiance $L_a/L_0$ (dimensionless); (e) $\log_{10}(L_0)$, being $L_0$ the zenith radiance in W·m$^{-2}$·sr$^{-1}$.

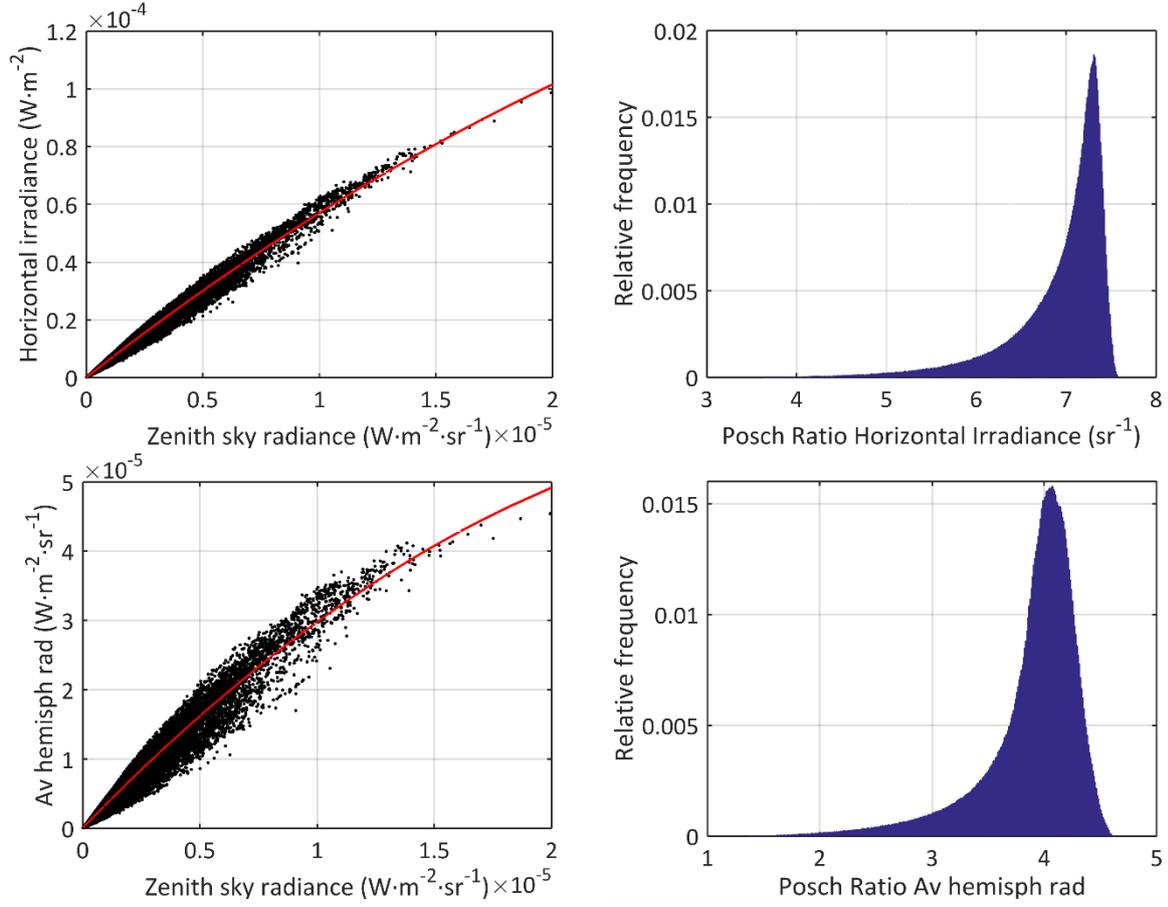

**Figure 5**. Top row: Scatter plots of the horizontal irradiance $E_0$ vs the zenith radiance $L_0$ (left), and histogram of the values of the Posch ratio for the horizontal irradiance, $E_0/L_0$ (right). Bottom row: idem, for the average upper hemisphere radiance $L_a$. Full, red lines in the scatter plots: fits of a polynomial of degree $n=2$ (see Fig 4, Table 1, and text for details).

**Table 1. Parameters of the fits in Figure 5 (left column),** $y = a_0 + a_1 x + a_2 x^2$

|  | **Horizontal irradiance** |  | **Average upper hemisphere radiance** |  |
| --- | --- | --- | --- | --- |
| **Parameter** | value (std error) | units | value (std error) | units |
| $a_0$ | 5.427e–08 (1.4e-10) *** | $W \cdot m^{-2}$ | 3.788e-08 (1.2e-10) *** | $W \cdot m^{-2} \cdot sr^{-1}$ |
| $a_1$ | 6.302e+00 (3.8e-04) *** | sr | 3.496e+00 (3.4e-04) *** | 1 |
| $a_2$ | –6.039e+04 (5.4e+01) *** | $W^{-1} \cdot m^2 \cdot sr^2$ | –5.192e+04 (4.8e+01) *** | $W^{-1} \cdot m^2 \cdot sr$ |
| RSE | 3.318e-07 | $W \cdot m^{-2}$ | 2.953e-07 | $W \cdot m^{-2} \cdot sr^{-1}$ |
| $R^2$ | 0.9913 | ---- | 0.9764 | ---- |

RSE: Residual standard error (on 7019603 degrees of freedom)
*** : p-value < 0.001

## 4. DISCUSSION

The value of the Posch ratio for any particular sky will be given by the weighted sum of the values of the individual Posch ratios that each type of source contributing to the sky radiance would produce separately. The weights of this linear combination are equal to the relative contributions of each source to the total zenith radiance. As shown in previous sections, the Posch ratio for the horizontal irradiance produced by the natural sky in sites free from light pollution during clear and moonless nights lies within the range of 1 to 1.7 times $\pi$ sr, being that way slightly higher than the one corresponding to a uniformly bright sky ($\pi$ sr). The light pollution component shows generally a larger Posch ratio, with a frequency histogram peaking at 2.25 times $\pi$ sr. The latter result is compatible with the well-known fact that the light polluted sky tends to be brighter towards the horizon, as expected from the atmospheric propagation of light produced by ground-level artificial sources. Notwithstanding that, the Posch ratio of the artificial irradiance diverges only moderately from its Lambertian reference value, because the contribution to the total irradiance of the largest radiances of the upper hemisphere, usually located close to the horizon rim, are weigthed by the cosine of the zenith angle and hence are substantially scaled down.

Among all considered radiance terms, moonlight is no doubt the one that shows a larger Posch ratio variability, reaching values remarkably larger than the ones associated with the other source types. This result can be traced back to the fact that the moonlit night sky has an extremely non-uniform brightness distribution, being the direct radiance of the Moon considerably larger than the scattered one. It is precisely that direct radiance the main responsible for the horizontal irradiance, excepting when the Moon is located at very low elevation angles. Notwithstanding their large range of variation, the Posch ratios associated with moonlight are deterministic and can be estimated if some basic information about the state of the atmosphere is known. They provide a practical way to estimate the expected value ranges of several useful radiance indicators, when measurements of the zenith radiance are the only information available.

Note that the Posch ratios associated with moonlight depend substantially on the Moon elevation angle but not so strongly on its phase, excepting for minor effects derived from the phase-dependent spectral composition of moonlight (Kieffer & Stone 2005) coupled with the differential wavelength scattering and absorption in the atmosphere. These effects are integrated within the photometric passband and are already included in the Johnson V band figures shown in subsection 3.2.

The results presented in this paper allow to interpret and to put in a broader context the observational data of the Posch ratio for the horizontal irradiance reported by Jechow et al (2020) after a campaign of measurements of the night sky radiance from urban to rural locations in the Berlin area. Based on all-sky photometry made with a commercial grade, calibrated DSLR camera fitted with a fisheye lens they found that despite the different characteristics of the measurement locations the Posch ratios were in the range from 1 to 2 times $\pi$ sr, values that are consistent with the expected ones in clear and moonless nights where both artificial sources and the natural starry sky contribute to build the overall sky brightness. Our results are also consistent with the ones previously reported by Kocifaj et al (2015), obtaining values of the horizontal irradiance to zenith radiance ratio in the range from 1.6 to 2 times $\pi$ sr for different canonical examples of artificial light propagation in exponentially stratified atmospheres.

The numerical results in section 3 were calculated for an unobstructed hemispheric field of view around the observer. In presence of dark obstacles (case of a very irregular topography, tall vegetation, or buildings nearby) the zenith sky brightness will be the same, but the linear radiance indicator in the numerator of the Posch ratio (e.g.: horizontal irradiance, average hemispheric radiance...) will generally be smaller due to the screening by the obstacles of the sky radiance at intermediate and low horizon angles. This will result in a PR smaller than the one shown in our examples. However, there is no fundamental difficulty in calculating the expected Posch ratios for any possible configuration of the obstacles within the observer's field of view. The general definitions and the procedures of calculation are largely the same as in the simplified cases presented in the examples above.

The Posch ratios concept and its formulation in the equations of this paper (including the appendix) are of general validity within the domain of linear systems theory and do not require any additional assumption about the possible symmetries that the point spread functions may or may not possess (e.g. translational or rotational invariance). These symmetries are mostly relevant for some practical purposes, e.g. to significantly speed up computationally demanding superposition integrals by resorting to Fourier domain techniques. The convenience of using simplified point spread functions instead of the exact ones shall be considered for each particular application, weighting the computational benefits against the potential accuracy losses.

## 5. CONCLUSIONS

The results presented in this work show that it is possible to get information about the expected values of several linear sky radiance indicators by measuring the value of one of them, e.g. the zenith sky radiance routinely recorded nowadays by several measurement networks worldwide. The ratios of the values of the indicators to the zenith radiance (Posch ratios) lie within specific ranges for each type of source contributing to the radiance of the night sky (starlight and other astropysical sources, moonlight and artificial light). A useful property of the Posch ratio is that its value for any particular sky is given by the linear combination of the ratios that each source would produce separately, the weights of this combination being the relative contribution of each source to the brightness of the zenith sky. Whereas the precise values of the individual Posch ratios depend on several atmospheric factors, their canonical value ranges for the examples described in this paper allow to get useful insights about what can be expected in any particular night. As such, they can be used to estimate the expected range of values of the horizontal irradiance or the average upper hemisphere radiance when the zenith radiance is known. The precision and accuracy of this estimation can be substantially improved if reliable information on the atmospheric conditions (AOD, aerosol types, etc) is available from real time measurements or can be established in a probabilistic way from historial datasets of measurements made at the desired location.

## ACKNOWLEDGEMENTS

This work was supported by Xunta de Galicia/FEDER, grant ED431B 2020/29 (SB). EM acknowledges partial funding by the Spanish MICIN/AEI/10.13039/501100011033 and by "ERDF A way of making Europe" by the "European Union" through grant RTI2018-095076-B-C21, and the Institute of Cosmos Sciences University of Barcelona (ICCUB, Unidad de Excelencia 'María de Maeztu') through grant CEX2019-000918-M. MK acknowledges funding from Slovak Research and Development Agency; contract No: APVV-18-0014.

## DATA AVAILABILITY

| Data available on request. | *The data underlying this article will be shared on reasonable request to the corresponding author.* |

**APPENDIX**

Let us denote by $L = L(\mathbf{r}, \boldsymbol{\alpha}, \lambda, t, \mathbf{p}) \equiv L(\mathbf{z})$ the spectral radiance of the sky, where $\mathbf{r}$ is the position vector of the observer, $\boldsymbol{\alpha} = (z, \phi)$ the direction of observation, being $z$ the zenith angle and $\phi$ the azimuth, $\lambda$ the wavelength, $t$ the time of observation, and $\mathbf{p}$ the polarization mode of the radiance field. A linear radiance indicator is defined as any physical quantity that can be obtained by the action of a suitably defined linear operator acting on several variables of the radiance, $B = \mathcal{L}[L]$. A particular subset of linear indicators are those defined by a weighted integral of the radiance. We will restrict the scope here to this particular subset, although the basic results shown below hold for any other family of linear indicators (e.g. differential ones like the gradient or the Laplacian of the radiance field).

Integral radiance indicators have the general form

$$B(\mathbf{x}) = \int_{\Xi} K(\mathbf{x}, \mathbf{y}) L(\mathbf{x}, \mathbf{y}) \mathrm{d}^n \mathbf{y} , \qquad (A.1)$$

where **x** is a vector composed of the subset of variables included in **z** on which the output indicator depends, **y** is a vector composed of the remaining $n$ variables included in **z** over which the integration is performed, $K(\mathbf{x}, \mathbf{y})$ is the operator kernel defining the indicator, and $\Xi$ is the vector space spanned by **y**.

The sky radiance $L(\mathbf{x}, \mathbf{y}) \equiv L(\mathbf{z})$, in turn, can be expressed in term of the sources' radiance (natural and/or artificial) as:

$$L(\mathbf{z}) = \int_H \Psi(\mathbf{z}, \mathbf{u}) L_s(\mathbf{u}) \mathrm{d}^m \mathbf{u} \quad (A.2)$$

where $L_s(\mathbf{u})$ is the radiance of the sources (depending on source position and angular emission variables **u**), the kernel $\Psi(\mathbf{z}, \mathbf{u})$ is the point spread function (PSF) for the sky radiance, and H is the $m$-dimensional space spanned by **u**.

An immediate result of Eqs. (A.1) and (A.2) is that the sky radiance indicators can be obtained by means of a suitable PSF acting directly on the sources' radiance, without the need of calculating explicity the value of the sky radiance itself, $L(\mathbf{z})$. By combining (A.1) and (A.2), we get

$$B(\mathbf{x}) = \int_\Xi K(\mathbf{x}, \mathbf{y}) \left[ \int_H \Psi(\mathbf{x}, \mathbf{y}, \mathbf{u}) L_s(\mathbf{u}) \mathrm{d}^m \mathbf{u} \right] \mathrm{d}^n \mathbf{y} = \int_H \left[ \int_\Xi K(\mathbf{x}, \mathbf{y}) \Psi(\mathbf{x}, \mathbf{y}, \mathbf{u}) \mathrm{d}^n \mathbf{y} \right] L_s(\mathbf{u}) \mathrm{d}^m \mathbf{u}$$

$$= \int_H \Gamma(\mathbf{x}, \mathbf{u}) L_s(\mathbf{u}) \mathrm{d}^m \mathbf{u} \quad (A.3)$$

where

$$\Gamma(\mathbf{x}, \mathbf{u}) = \int_\Xi K(\mathbf{x}, \mathbf{y}) \Psi(\mathbf{x}, \mathbf{y}, \mathbf{u}) \mathrm{d}^n \mathbf{y} \quad (A.4)$$

is the point spread function that allows calculating $B(\mathbf{x})$ as a direct integral over the sources $L_s(\mathbf{u})$ (Bará & Lima (2018); Bará et al 2021; Falchi & Bará 2021).

The values of any pair of linear sky radiance indicators are expected to be correlated, depending on the degree of correlation of the source emissions:

$$\langle B_1(\mathbf{x_1}) B_2(\mathbf{x_2}) \rangle = \langle \int_H \Gamma_1(\mathbf{x_1}, \mathbf{u}) L_s(\mathbf{u}) \mathrm{d}^m \mathbf{u} \int_{H'} \Gamma_2(\mathbf{x_2}, \mathbf{u}') L_s(\mathbf{u}') \mathrm{d}^m \mathbf{u}' \rangle$$

$$= \int_H \int_{H'} \Gamma_1(\mathbf{x_1}, \mathbf{u}) \Gamma_2(\mathbf{x_2}, \mathbf{u}') \langle L_s(\mathbf{u}) L_s(\mathbf{u}') \rangle \mathrm{d}^m \mathbf{u} \, \mathrm{d}^m \mathbf{u}' \quad (A.5)$$

where $\langle \cdot \rangle$ indicates averaging over the domains of the relevant variables. In general, even for a random, totally uncorrelated spatial source distribution, $\langle L_s(\mathbf{u}) L_s(\mathbf{u}') \rangle = L_s^2(\mathbf{u})\delta(\mathbf{u} - \mathbf{u}')$, some degree of correlation it is expected to exist between the indicators, due to the mixing effects of the PSFs:

$$\langle B_1(\mathbf{x_1})B_2(\mathbf{x_2}) \rangle = \int_H \Gamma_1(\mathbf{x_1}, \mathbf{u}) \, \Gamma_2(\mathbf{x_2}, \mathbf{u}) \, L_s^2(\mathbf{u}) \, d^m\mathbf{u} \qquad (A.6)$$

The Posch ratio for any pair of linear sky radiance indicators, $B_1(\mathbf{x_1})$ and $B_2(\mathbf{x_2})$, is defined as:

$$\Pi_{1,2}(\mathbf{x_1}, \mathbf{x_2}) \equiv \frac{B_1(\mathbf{x_1})}{B_2(\mathbf{x_2})} \qquad (A.7)$$

so that

$$\Pi_{1,2}(\mathbf{x_1}, \mathbf{x_2}) = \frac{\int_H \Gamma_1(\mathbf{x_1}, \mathbf{u})L_s(\mathbf{u})d^m\mathbf{u}}{\int_H \Gamma_2(\mathbf{x_2}, \mathbf{u})L_s(\mathbf{u})d^m\mathbf{u}} \qquad (A.8)$$

Recalling that the Posch ratio of the values of the indicators produced by an elementary source located at $\mathbf{u}$ is by definition $\Pi_{1,2}(\mathbf{x_1}, \mathbf{x_2}; \mathbf{u}) = \Gamma_1(\mathbf{x_1}, \mathbf{u})/\Gamma_2(\mathbf{x_2}, \mathbf{u})$ (the $L_s(\mathbf{u})d^m\mathbf{u}$ factors cancel out) we can rewrite eq (A.8) as:

$$\Pi_{1,2}(\mathbf{x_1}, \mathbf{x_2}) = \frac{\int_H \Gamma_2(\mathbf{x_2}, \mathbf{u})\Pi_{1,2}(\mathbf{x_1}, \mathbf{x_2}; \mathbf{u})L_s(\mathbf{u})d^m\mathbf{u}}{\int_H \Gamma_2(\mathbf{x_2}, \mathbf{u})L_s(\mathbf{u})d^m\mathbf{u}} = \int_H \gamma(\mathbf{x_2}; \mathbf{u}) \, \Pi_{1,2}(\mathbf{x_1}, \mathbf{x_2}; \mathbf{u}) \, d^m\mathbf{u} \qquad (A.9)$$

where $\gamma(\mathbf{x_2}; \mathbf{u})$ is the fractional contribution of the sources located at $\mathbf{u}$ -per unit element of $d^m\mathbf{u}$- to the total value of the indicator $B_2(\mathbf{x_2})$

$$\gamma(\mathbf{x_2}; \mathbf{u}) = \frac{\Gamma_2(\mathbf{x_2}, \mathbf{u})L_s(\mathbf{u})}{\int_H \Gamma_2(\mathbf{x_2}, \mathbf{u}')L_s(\mathbf{u}')d^m\mathbf{u}'} \qquad (A.10)$$